\documentclass [aip,jmp, amsmath,amssymb,reprint]{revtex4-1}

\usepackage{amssymb}
\usepackage{amsbsy}
\usepackage{amsmath}
\usepackage{amsfonts}
\usepackage{mathrsfs}
\usepackage{latexsym}
\usepackage[english]{babel}
\usepackage{url}
\usepackage[]{graphicx}
\usepackage{subfigure}
\usepackage{hyperref}
\begin{document}

\title{Vortex-crystal pattern in an anti-artificial spin ice system}

\author{C.I.L. de Araujo}
\email{dearaujo@ufv.br} 
\affiliation{ Departamento de F\'{i}sica, Universidade Federal de Vi\c cosa, Vi\c cosa 36570-900, Minas Gerais, Brazil }

\author{R. C. Silva}\affiliation{  Departamento de F\'{i}sica,
Universidade Federal de Vi\c cosa, Vi\c cosa 36570-900, Minas Gerais, Brazil }

\author{I. R. B. Ribeiro}
\affiliation{  Departamento de F\'{i}sica,
Universidade Federal de Vi\c cosa, Vi\c cosa 36570-900, Minas Gerais, Brazil }

\author{F.S. Nascimento}
\affiliation{  Departamento de F\'{i}sica,
Universidade Federal de Vi\c cosa, Vi\c cosa 36570-900, Minas Gerais, Brazil }

\author{J.F. Felix}\affiliation{ Departamento de F\'{i}sica,
Universidade Federal de Vi\c cosa, Vi\c cosa 36570-900, Minas Gerais, Brazil }

\author{S.O. Ferreira}\affiliation{ Departamento de F\'{i}sica,
Universidade Federal de Vi\c cosa, Vi\c cosa 36570-900, Minas Gerais, Brazil }
\author{L.A.S. M\'{o}l}\affiliation{Departamento de F\'{i}sica, ICEx, Universidade Federal de Minas Gerais, Belo Horizonte 31270-901, Minas Gerais, Brazil.}

\author{W.\ A.\  Moura-Melo}
\affiliation{ Departamento de F\'{i}sica,
Universidade Federal de Vi\c cosa, Vi\c cosa 36570-900, Minas Gerais, Brazil }

\author{A.\ R.\  Pereira}
\affiliation{ Departamento de F\'{i}sica,
Universidade Federal de Vi\c cosa, Vi\c cosa 36570-900, Minas Gerais, Brazil }

\date{\today}

\begin{abstract}

We have proposed in this work an original system composed by anti-dots nanopatterned in a ferromagnetic thin film, mimicking negatively the structure of an articial spin ice. In the hysteresis loop we notice the emergency of an anisotropy in the magnetization saturation and in the micromagnetic simulations, in the beginning of the hysteresis loop (relaxation), the formation of a vortex crystal array with vortices in diferent positions possessing random polarization and chirality. The crystal of vortices in this electrically connected sample could be most eficient than those observed in non-connected nanodiscs for current-driven or magnetic vortices switching by electric currents. 

\end{abstract}

\keywords{Vortex, anti-dots, spin-ice, micromagnetics, spintronics}

\maketitle

Nowadays, nanomagnetism is one of the most exciting branches of basic and applied research. Several applications have become reality and many others are under way. The experimental advances to fabricate and characterize nanosamples have provided a lot of new magnetic phenomena and possibilities at such scale. For instance, magnetic thin films, with thickness around nanometers, display peculiar magnetization patterns depending on its global shape: a thin nanodisk can support a single topological vortex-like configuration, while in a square (rectangular) shape, multi-vortex (Landau) states show up. On the other hand, elongated shapes, like cylinders and ellipsoids, generally display single-domain patterns with strong Ising-like anisotropy along the major dimension (see Refs.~\onlinecite{review1,review2} for reviews). The influence of holes (antidots) intentionally introduced in a magnetic nanodisk on the vortex structure and dynamics has also been intensively investigated \cite{Rahm03,mag-logic-storage, Pereira05,Ricardo08}. Many applications of those topological objects were raised, including magnetic logic and storage\cite{mag-logic-storage} with promising technological application in spintronics, magneto-mechanical apoptosis reactivation in cancer cells\cite{nanodisk-cancer}, among many others.
\begin{figure}[!h]
		\center
		\includegraphics[width=8.0cm]{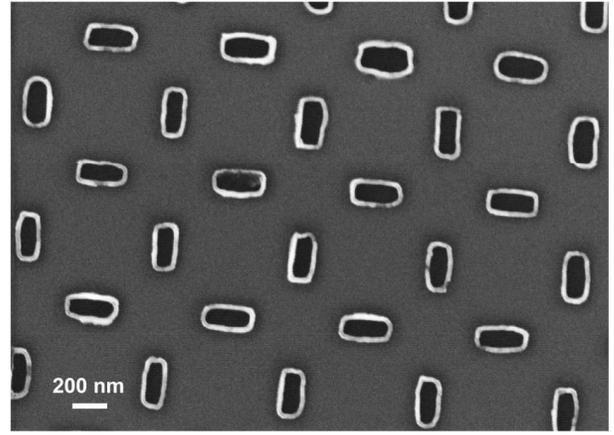}
		\caption{SEM-FEG micrography of the anti-artificial spin ice sample composed by $25 nm$ nickel thin film (grey) and black holes mimicking the spin ice geometry.}
		\label{fig:fig1}
	\end{figure}	
Elongated ferromagnetic nanoislands displayed in suitable two-dimensional ($2D$) arrangements have also attracted a lot of attention, yielding frameworks resembling water ice. Actually, in such artificially structured arrangements, geometrical frustration prevents them to be completely frozen, so that they bear residual entropy even at zero temperature. For this resemblance with water ice, those systems have been named artificial spin ices (ASI) \cite{Wang2006}. Among several interesting properties, these systems were predicted \cite{Mol09,Mol10,Silva12} to support emergent magnetic monopole excitations interacting via a Coulomb potential, similarly to what had already been observed in three-dimensional ($3D$) spin ice crystals\cite{Ryzhkin05,Castelnovo08,Bramwell09,Morris09,Kadowaski09}; however, in the $2D$ ASI, the oppositive poles are connected by an additional energetic string\cite{Silva12} (leading to a kind of a pair of Nambu monopoles \cite{Nambu}). These predictions were soon confirmed by experiments\cite{Morgan11}. Such findings shed some light into the real possibilities of using isolated magnetic poles as the physical carriers of charge, energy and momentum - a kind of electricity with magnetic charges, the so called magnetricity. In order to diminish the string tension and provide monopoles separation inside artificial spin ice samples, different geometries and insertion of defects, like non-uniformity or dislocation of the cell sizes, have been proposed \cite{rodrigo}. It should be also remarked the timely interest in nanometric anti-dots intentionally inserted in continuous thin films, in which they are disposed in regular geometrical arrangements depicting a number of patterns, as diamond-type, square, triangular, and so forth \cite{Wiele,Tse,Luo}. The main results of these works are the changes observed in the demagnetization field and domain wall pinning, which bring about the emergence of magnetic anisotropy in the patterned thin films as well as a huge increasing in the coercive field. Such a behavior strongly suggests anti-dots systems as promising structures for application in high-density magnetic storage, predicted to be around ($0.1 Tb cm^{2}$) and reading/writing velocity of ($0.5 Gb s^{-1}$) \cite{Torres,Weller}.

Here, we report on the experimental fabrication and theoretical study of a novel structure composed by a lattice of anti-dots, negatively mimicking the spin ice geometries in a continuous thin film. This system was built as follows. On the silicon ($110$) substrate with area of $1 cm^{2}$ previously cleaned by $RCA$ process, a $200 nm$ Polymethylmethacrylate ($PMMA$) layer was deposited by spin coating at $4000 rpm$ for $40$ seconds. The $PMMA$, dried for $10 min$ in hotplate at $120^{o}C$, has $1nm$ $RMS$ roughness measured by atomic force microscopy. The sample was carried into the $RAITH$ $e_{-} LINE$  plus system chamber, where the exposure of the anti-spin ice design was performed. Such design was repeated a thousand times to form $1mm^{2}$ lithographed area. The parameters utilized were beam acceleration of $20KV$ and area cleaning dose of $160 C/cm^{2}$. After the $PMMA$ developing, the samples were placed into the Thermionics $E_{-} Beam$ evaporation system where a $25nm$ nickel film was evaporated 
over a $8nm$ titanium seed layer, used to increase the adhesion. A cap layer of $3nm$ gold was evaporated over the sample to prevent the nickel oxidation. Finally the structures were defined by lift-off process in acetone ultrasonic bath. The final samples structures where investigated by Scanning Electron Microscopy with Field Emission Gun ($SEM-FEG$). In order to investigate the influence of our patterned anti-ASI on the thin film hysteresis loop, magnetic characterizations were realized in a Microsense $EV9$ Vibrating Sample Magnetometer ($VSM$). The total magnetization of the samples were taken in configurations with external magnetic field $\vec{H}$ applied at $0^{o}$ and $45^{o}$ in relation to the anti-ASI side. In the Figure  \ref{fig:fig1} we present the Scanning Electronic Microscopy (SEM-FEG) image of the final sample where the nickel thin film is represented by the grey color and the anti-ASI by the black elongated holes.
\begin{figure}[hbt]
		\centering
		\includegraphics[width=8.0cm]{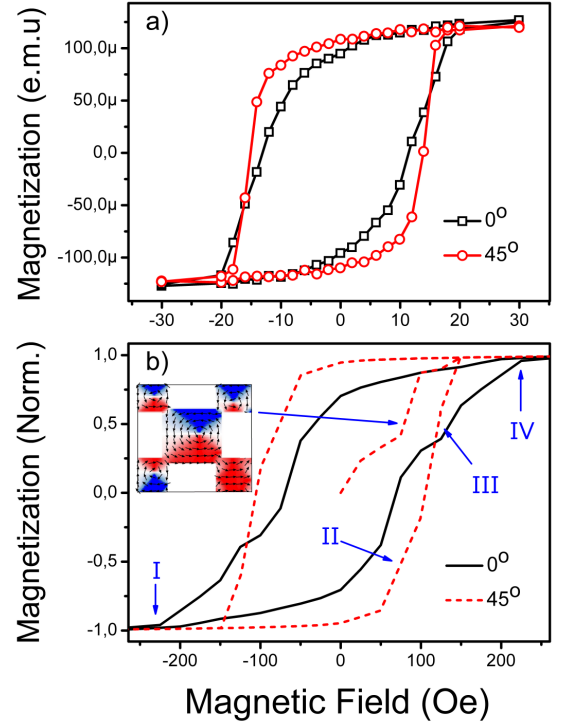}\\
		\caption{a) Magnetization curves obtained by Vibrating Sample Magnetometer ($VSM$) with field oriented at $0^{o}$ and $45^{o}$ in respect to the anti-ASI sample. b) hysteresis loop from the micromagnetic simulation.
 }
		\label{fig:fig2}
	\end{figure}

From the $VSM$ measurements, presented in Figure 2a, it is possible to note the squareness in the hysteresis loop and an increase on the coercive field when the external magnetic field is applied in a diagonal direction $45^{o}$ in relation to the configuration with the field applied along the square sides $0^{o}$.

Below, we theoretically describe the main characteristics of the system which should be responsible for the observed behaviors of the hysteresis. In order to analyze spin configurations in different hysteresis curve regions at $0^{o}$ and $45^{o}$,  micromagnetic simulation was performed with computational codes provided by the Object Oriented MicroMagnetic Framework ($OOMMF$) \cite{oommf}, which is based upon the Landau-Lifshitz-Gilbert ($LLG$) equation \cite{landau,gilbert} and effectively describes magnetization $\vec{M}$ dynamics:

\begin{equation}
\frac{\partial \vec{M}}{\partial t}= -\gamma \vec{M} \times \vec{H}_{eff} + \frac{\alpha}{M_{s}}\vec{M} \times \frac{\partial \vec{M}}{\partial t}\,,
\end{equation}
where $\gamma$ is the gyromagnetic ratio, $M_{s}$ the saturation magnetization, while $H_{eff}$ accounts for the effective magnetic field  (which is composed by external magnetic field, magneto-crystalline anisotropies, dipolar and exchange interactions). Such an equation is used to determine the minimum energy and the transitions between spin configurations. The simulation parameters adopted for nickel $Ni$ are: exchange (stiffness) constant $A^{Ni}=9 \times 10^{-12}  J/m$ and saturation magnetization $M_{s}=4.3 \times 10^{5}  A/m$ . The structures were divided in meshes with sizes of $5nm^{3}$, which are smaller than nickel exchange length, $l_{\rm ex}=\sqrt{A/4\pi M^2_s}\approx 7.72 {\rm nm}$. The hysteresis loop obtained by simulation is shown in Figure 2b and it is in good qualitative agreement with our experimental measurements.{\bf}

Since our simulations have been performed taking into account a small portion of the whole system (around $3 \times 3 \mu m^{2}$, for saving computational efforts), the theoretical field range is much higher than that of the experimental results. It is an indication that the whole sample should be patterned for possible applications in magnetic storage. In our case, only $1mm^{2}$  from the $1cm^{2}$ of the sample was patterned and so the experimental hysteresis loop has a huge influence of the large area of non-patterned nickel thin film.
\begin{figure}[hbt]
		\centering
		\includegraphics[width=8.2cm]{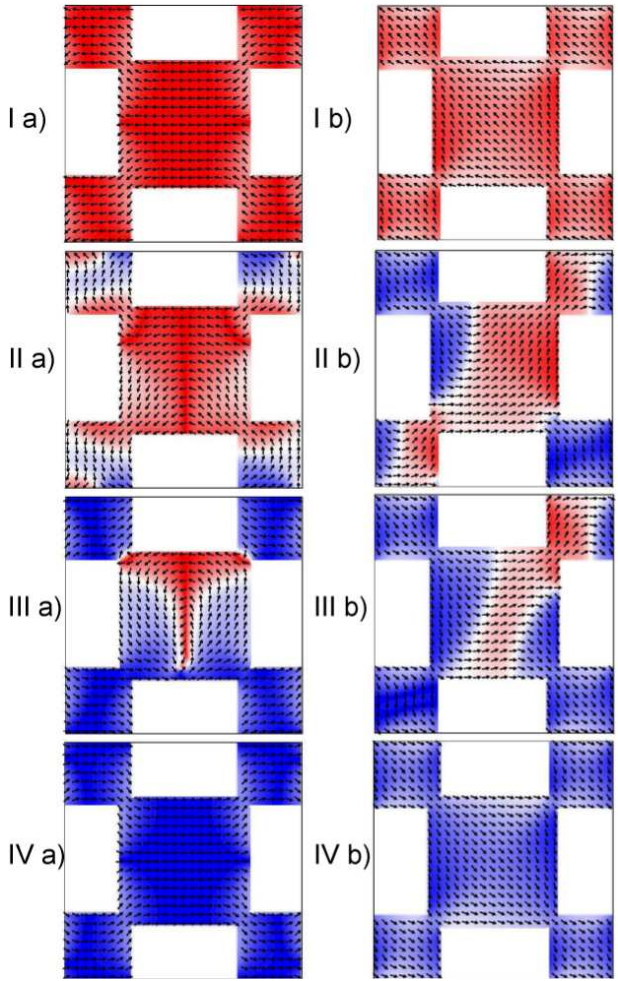}\\
		\caption{Micromagnetic analyze of domain wall dynamics in one cell made during the hysteresis simulation (\ref{fig:fig2}), for a) the applied magnetic field is longitudinal to the cell $0^{o}$ and for b) the field is diagonal to the cell $45^{o}$. The red and blue areas represent the spins saturated to negative and positive fields respectively.  }
				\label{fig:fig3}
	\end{figure}
From the magnetic configuration in each range of the hysteresis loop obtained by simulations (see Fig.  \ref{fig:fig3}), it is possible to note that the spins are parallel to the system's border in order to reduce the magnetostatic energy. So, in the square geometry, when the magnetic field is applied along the sides of the cells (Figure 3a I-IV), the influence of the pinned domains (in the anti-ASI cells sides) perpendicular to the field has weak contribution to the pinning process and a high field is necessary for the total magnetic saturation. When the field is applied along the diagonal of the cells (Figure 3b I-IV) both pinned domains contribute to the squareness and increase in the coercivity of the hysteresis loop; this behaviour corroborates our experimental observation.

\begin{figure}[!h]
		\centering
		\includegraphics[width=6.0cm]{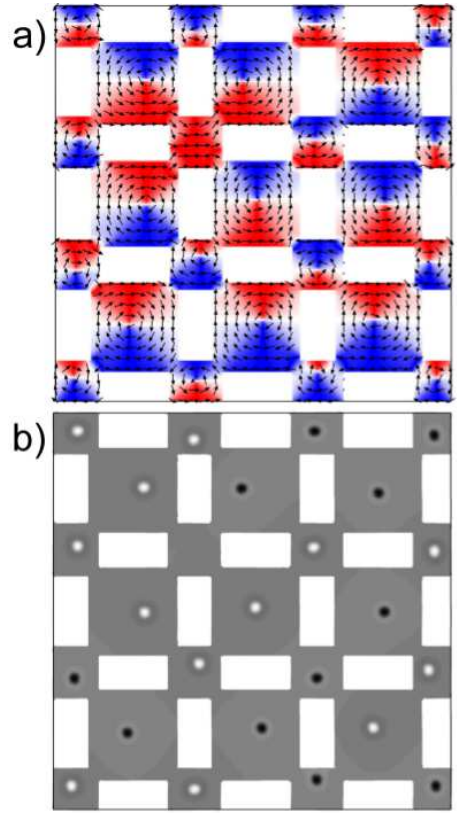}\\
				\caption{a) Simulation of $9$ cells showing the crystal of vortices in random chirality. b) Core polarization: black and white spots represent oppositive directions}
				\label{fig:fig4}
	\end{figure}

Whenever analysing the magnetization behavior during spins relaxation in the beginning of the hysteresis curve (see the inset in Figure 2b), we have realized the appearance of vortex-crystal patterns throughout the sample. The vortex-crystal is presented in the Figure 4a and the vortices chirality follows a random distribution. Our simulations show that the vortex-crystal appears only when the $Ni$ thickness is above $25nm$. As the expected Curie temperature $T_{c}$ in this thickness is close to the bulk value of $631K$ \cite{sun}, this vortex-crystal could be observed well above room temperature. The spins near the vortex centers become out of plane (up or down polarization) in order to minimize the magnetostatic energy. Figure 4b illustrates the polarizations of the vortex-crystal as obtained in the simulations. The black and white dots represent opposite core polarizations and our results indicate that the vortex polarizations over the crystal also follow a random distribution. Koyama \textit{et al}. have shown the possibility of aligning every vortex in an array of nanodiscs with the same polarity and chirality by applying constant magnetic field perpendicular to the sample \cite{koyama}; Kasai $et$ $al.$ demonstrated (by simulation and experimental realization) the vortices switch by application of $ac$ electric current \cite{kasai}.

In summary, we have nanofabricated an anti-ASI on nickel thin film. Our experimental results show an increase in the coercivity and squareness of the hysteresis loop and its behavior was confirmed by micromagnetic simulations. Previously, similar results have already been observed in other anti-dot systems \cite{kohli,wang}. One of the most important result presented here is the role of the proposed geometry in the formation of a topological crystal of vortices in the ferromagnetic thin film. These vortices may induce several interesting properties in the system. For instance, the possibility of creation, manipulation and maintenance of such vortex-crystal in an electrically connected sample at high temperature could be very useful for future applications in spintronics.

The authors are grateful to P.R. Soledade and J.P. Sinnecker (LABNANO/CBPF-Brazil) for technical support during electron microscopy/nanolithography work and to Prof. A.A. Pasa (UFSC-Brazil), for the utilization of LFFS and LMCMM facilities. They also thank CAPES, CNPq and FAPEMIG (Brazilian agencies) for partial financial support.



\end{document}